\documentclass[prl, twocolumn,showpacs,preprintnumbers,amsmath,amssymb]{revtex4-1}
\usepackage{graphicx}
\usepackage{bm}
\newcommand{\unit}[1]{\ensuremath{\, \mathrm{#1}}}

\begin{document}
\title{High Repetition-Rate Wakefield Electron Source Generated by Few-millijoule, 30 femtosecond Laser Pulses on a Density Downramp}

\author{Z.-H.~He}
\author{B. Hou}
\author{J.~H.~Easter}
\author{K.~Krushelnick}
\author{J.~A.~Nees}
\author{A.~G.~R.~Thomas}

\affiliation{Center for Ultrafast Optical Science, University of Michigan, Ann Arbor, MI 48109-2099 USA}

\pacs{41.75.Jv, 52.38.Kd, 52.65.Rr, 29.25.Bx}

\begin{abstract}
We report on an experimental demonstration of laser wakefield electron acceleration using a sub-TW power laser by  tightly focusing 30-fs laser pulses with only $8\unit{mJ}$ pulse energy on a 100 $\mu$m scale gas target. The experiments are carried out at an unprecedented $0.5\unit{kHz}$ repetition rate,  allowing ``real time" optimization of accelerator parameters. Well-collimated and stable electron beams with a quasi-monoenergetic peak in excess of $100\unit{keV}$ are  measured. Particle-in-cell simulations show excellent agreement with the experimental results and suggest an acceleration mechanism based on electron trapping on the density downramp, due to the time varying phase velocity of the plasma waves. 
\\
\end{abstract} 
\maketitle

Since the concept of laser driven plasma accelerators was first proposed by Tajima and Dawson \cite{Tajima_PRL_1979}, advances in high-power ultrafast laser technology have enabled successful production of energetic electron beams in numerous wakefield acceleration experiments\cite{Mangles_Nature_2004,Geddes_Nature_2004,Faure_Nature_2004}. Plasma-based particle acceleration holds significant promise for future compact sources of relativistic electron beams because of the large acceleration gradients plasma can sustain relative to conventional radio frequency cavities. A high intensity laser pulse propagating in an underdense plasma generates large amplitude plasma waves with phase velocities close to the speed of light. Under certain conditions, electrons can be trapped in the waves and accelerated to relativistic energies. Recent progress has demonstrated that using ultrashort laser pulses, relativistic electron beams up to GeV energies with low energy spread and small divergence can be produced in a highly nonlinear yet stable mechanism known as the ``bubble'' regime \cite{Pukhov_APB_2002}. 

Accelerating electrons in the ``bubble'' regime requires a laser pulse that is both intense (with $a_0>1$, where $a_0=eA/m_ec$ is the normalized vector potential) and short (with pulse duration $\tau\leq2\pi c/\omega_p$, 
 where $\omega_p=\sqrt{e^2n_e/m_e\epsilon_0}$ is the plasma frequency). In typical experiments, the intense laser pulse is focused onto the edge of a supersonic gas jet with a matched spot size. The densities of such gaseous targets indicate that ultrafast laser systems with pulse energies on the order of a joule or more are necessary. Earlier experiments using longer laser pulses also accelerated electrons via a self-modulation instability where the laser pulse length was much greater than the  wavelength of a relativistic plasma wave, $\tau>2\pi c/\omega_p$
\cite{Joshi_PRL_1981,Andreev_PPR_1995,Nakajima_PRL_1995,Modena_Nature_1995,Ting_POP_1997}. However, these experiments were limited to operation at a low repetition rate ($<1$ Hz) due to the high laser pulse energies.  

Using a low energy laser pulse to accelerate electrons requires a short underdense plasma, as laser depletion in generating the plasma wave limits the acceleration length. Conditions for electron trapping in the plasma wave are also restrictive due to the lower achievable intensity. One method for trapping electrons is to use a density downramp injection mechanism \cite{Bulanov_PRE_1998}, which was recently demonstrated experimentally using 10 TW lasers \cite{Geddes_PRL_2008,Schmid_PRSTA_2010}. In this scheme, the inhomogenous plasma leads to a time varying plasma wave phase velocity, which allows trapping once the electron velocity $v_e$ exceeds the wave phase velocity $v_{ph}$, $v_e>v_{ph}$.

In this Letter, we report on electron acceleration in an unexplored regime of plasma wakefield driven by few-millijoule femtosecond laser pulses (sub-TW) at high repetition rate (0.5 kHz). The high repetition rate enables better statistics and higher average particle flux, which were not accessible in previous experiments. Collimated electron beams are produced with quasi-monoenergetic spectra up to approximately 150 keV by acceleration in slow (non-relativistic) plasma waves on the density downramp of a 100 $\mu$m scale gas target. Because of the relatively high charge ($\sim10\unit{fC}$) and potentially short temporal duration, such electron sources have the potential to be used for ultrafast electron diffraction (UED) applications. 

In conventional UED, electrons from femtosecond laser pulse induced photoemission are accelerated by an external electric field. It is a challenging problem to control the amount of charge and the temporal resolution of the electron bunch. Laser acceleration eliminates external acceleration instruments and stabilizes the electron bunch charge and duration. Recently, laser-accelerated electrons from solid target interactions have been demonstrated to successfully produce diffraction patterns \cite{Tokita_APL_2009}. Additional energy selection using bending magnets was crucial in Ref.~\cite{Tokita_APL_2009} for obtaining mono-energetic electron pulses because the laser-accelerated electrons from solid targets have a very broad energy spectrum. In our experiments, the observed quasi-monoenergetic feature at 100 keV can be exploited to allow a larger portion of electrons to be available for applications in electron diffraction and microscopy. The gas target also permits operation at higher repetition rate with easier alignment.

The experiments were performed using the $\lambda^3$ laser system at the Center for Ultrafast Optical Science of the University of Michigan. This Ti:Sapphire based chirped pulse amplification (CPA) laser has a regenerative amplifier and a two-pass amplifier. It is capable of delivering pulses with energies up to 10 mJ and durations of 30 fs at a central wavelength of 800 nm. The output laser pulse is reflected from a deformable mirror (DM) and focused by either an $f/2$ or $f/3$ off-axis paraboloidal mirror to an optimized spot. Up to 8 mJ of pulse energy is available on target, which produces a peak intensity of $3\times10^{18}$ W/cm$^2$ in a 2.5 $\mu$m (FWHM) focal spot. The laser focus is optimized by iteratively setting the DM so that the signal of second-harmonic generation from a BBO crystal is maximized. 

\begin{figure}[htbp]
\begin{center}
\includegraphics[width=0.5\textwidth]{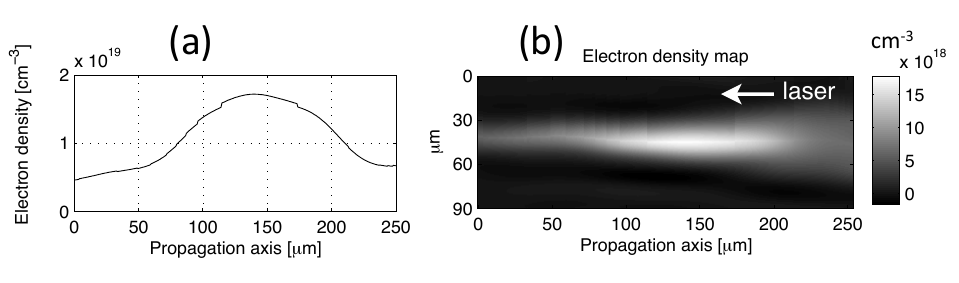}
\caption{(a) Lineout of electron on-axis density. (b) Reconstructed electron density map. }
\label{density}
\end{center}
\end{figure}

To achieve high repetition rates with the desired small diameter gas profile, we used a free flowing capillary source. The gas target was produced by flowing pure argon or helium with a 5\% impurity of nitrogen through the fused silica capillary tubing having an inner diameter of 100 $\mu$m. An approximately 10 cm length of this tubing was connected to a standard compressed gas system. A motorized XYZ stage was used to  manipulate the capillary tubing to an accuracy of 2 $\mu$m. The gas flow experiences free expansion into vacuum and different densities were achieved by varying the backing pressure. A Mach-Zehnder interferometer configuration, using a beam containing 2\% energy split from the main beam and probing the gas flow transversely, was used to measure the plasma density profile. The interferogram data provide diagnostics on both the plasma spatial distribution and its temporal evolution, by varying the probe delay time. A 2D electron density map was reconstructed via Abel inversion of the phase-shift data as shown in Fig.~\ref{density}. 

To measure the laser-accelerated electron beam profile, a Lanex scintillating screen was placed approximately 8 cm from the nozzle along the laser propagation axis. The Lanex screen was covered by a $17\pm2\unit{\mu m}$ thick aluminum foil to block the laser light and cut off electrons with energies below $\sim50\unit{keV}$. Electron energies were measured by horizontal magnetic deflection from their original propagation path and imaging on the Lanex screen. An insertable pair of disc magnets yoked together providing a maximum magnetic field strength of 0.02 T at the midplane was used for this purpose. A $0.5\unit{mm}$ aluminum slit was placed vertically before the magnets. The electron energy distribution was also measured with a custom built magnet spectrometer \cite{Aghapi_PRL_2009} having an entrance aperture with a diameter of 3 mm and a solid angle acceptance of 1 msr. The electron spectra were recorded on image plates (FUJI BAS-SR 2025, calibrated in \cite{Tanaka_RSI_2005}).

\begin{figure}[htbp]
\begin{center}
\includegraphics[width=0.35\textwidth]{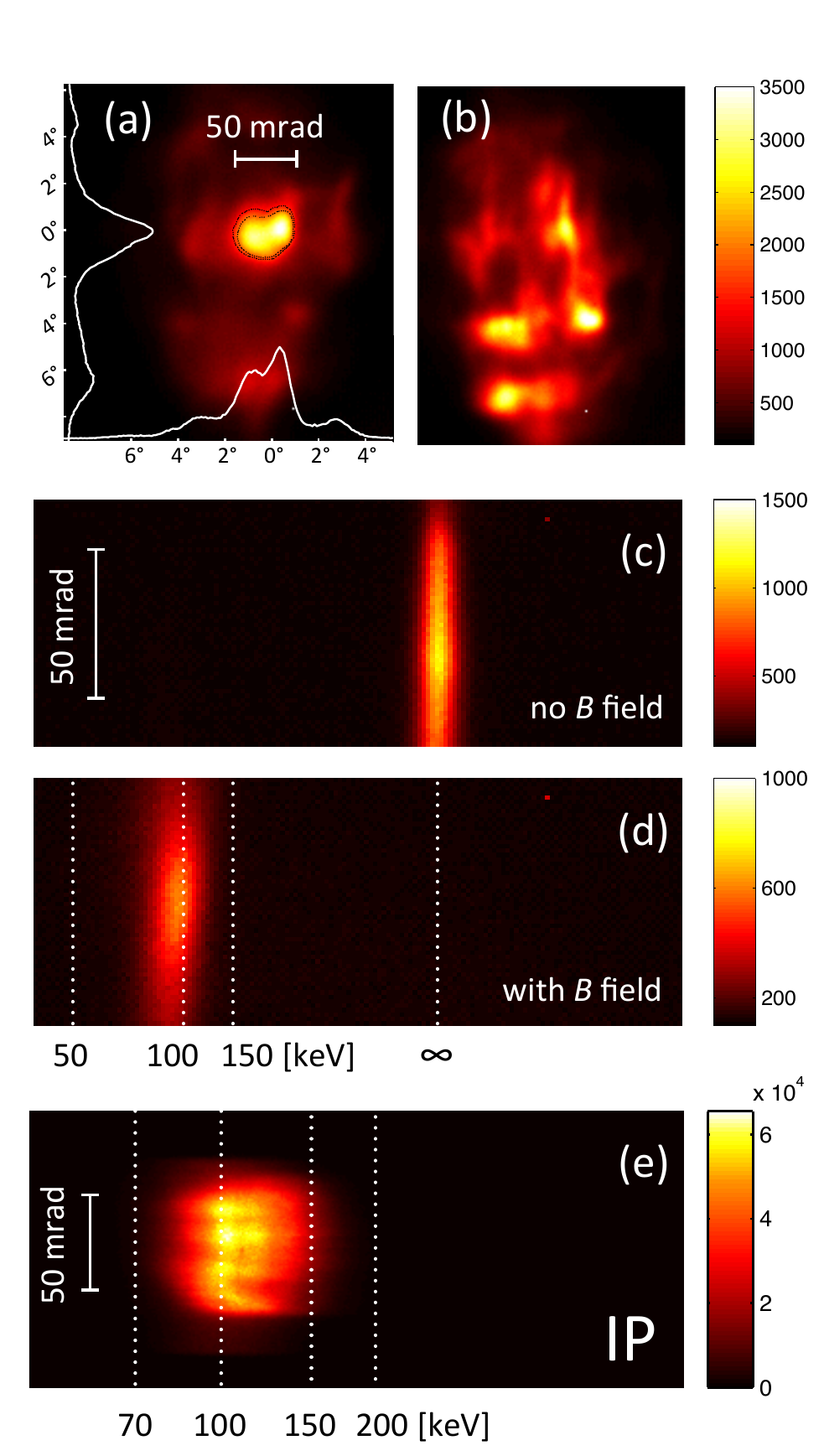}
\caption{(color online). (a)-(b): Images of electron beam profiles measured on Lanex screen (He:N$_2$ mixture gas. Exposure time $0.5\unit{s}$) (a) a typical collimated electron beam. The white solid curves are vertical and horizontal lineouts. The dotted contours (half-maximum) show the maximum fluctuation over six data acquisition sequences; (b) filamented electron beam. (c)-(d): Lanex images showing the electrons deflected by the insertable magnets using He:N$_2$ mixture (c) without magnets (d) with magnets. (e) Image plate raw data obtained using argon with the electron spectrometer accumulated over 2000 shots.} 
\label{fig1}
\end{center}
\end{figure}
Typical electron beam profiles are shown in Fig. \ref{fig1}(a) and (b). Well-collimated electron beams with a divergence angle (FWHM) as small as 2$^\circ$ were observed at lower densities [Fig. \ref{fig1}(a)]. As the backing pressure was increased and the nozzle position was lowered, the electron beams broke up into several filaments [Fig. \ref{fig1}(b)]. When the laser was focused at a larger distance above the nozzle, the plasma length became greater due to gas expansion and the peak plasma density decreased, which was confirmed by plasma interferometry. The overall effect could be a longer interaction length at similar plasma densities leading to electron beam filamentation, as was also observed in laser wakefield experiments on 30 terawatt laser systems \cite{Huntington_PRL_2011}. In both cases in these experiments, the features of the beam profiles remain stable when the system operated at $0.5\unit{kHz}$. We also used an on-axis silicon PIN diode detector covered by aluminum foil to measure the shot-to-shot stability of the electron signal. The oscilloscope trace from the diode consistently showed less than 10\%  fluctuation.

\begin{figure}[htbp]
\begin{center}
\includegraphics[width=0.5\textwidth]{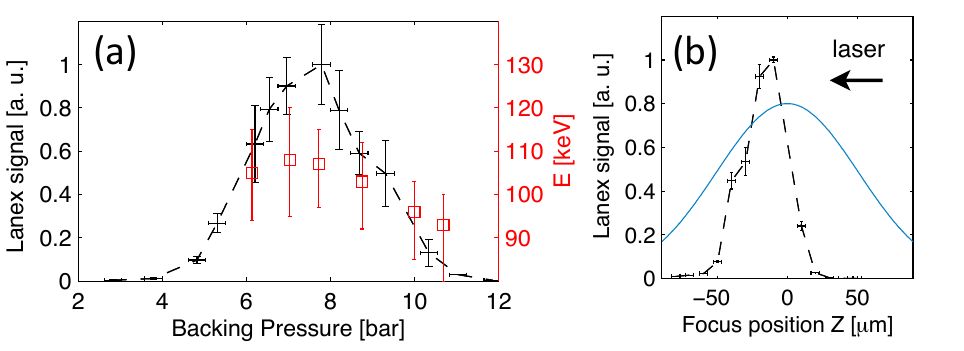}
\caption{(color online). (a) Measured electron beam charge and peak energy over a range of backing pressures. (b) Electron charge measured by Lanex signal versus laser focus position relative to the center of the capillary nozzle. $Z<0$ means laser is focused on the rear side of the nozzle. The blue line is the shape of a Gaussian density profile with a FWHM of $120\unit{\mu m}$.}
\label{scan}
\end{center}
\end{figure}

The raw images of the spectrum measurement are shown in Fig.~\ref{fig1}(c)-(e) with calibration lines. The measured spectrum exhibits a quasi-monoenergetic peak in the 100 keV energy range with width $\Delta E_{\unit{FWHM}}\approx20$ keV. In the experiments, data were obtained in ``real time" for optimization of the beam parameters. For a fixed nozzle position, electrons were observed over a finite range of backing pressures [Fig.~\ref{scan}(a)] corresponding to plasma densities on the order of $10^{19}$~cm$^{-3}$,  inferred from interferometric measurement (Fig.~\ref{density}). At lower densities, the plasma wave phase velocity is so high that the oscillating electrons are below the trapping threshold. At plasma densities where the plasma wake wavelength $\lambda_p$ is comparable to the laser pulse length $L$, large amplitude plasma waves are ``resonantly'' excited, enabling acceleration of fast electrons. 

At higher densities, the laser pulse is susceptible to plasma defocusing or filamentation instability. The Lanex signal showed that electrons were preferably produced when the laser was focused on the rear side of the nozzle [see Fig.~\ref{scan}(b)]. This is likely to be related to the acceleration mechanism based on density downramp injection \cite{Geddes_PRL_2008}, which will be discussed later. In the experiments, a gas mixture of helium and nitrogen was used under the consideration that electron trapping may be enhanced by additional ionization of the inner shell electrons near the peak laser intensity from a higher atomic number gas (e.g. nitrogen) \cite{Ionization}. However, the standard ionization induced trapping is not observed in our experiments, although the gas mixture yields a higher electron density compared to a pure helium gas with the same gas profile.

\begin{figure}[hbp]
\begin{center}
\includegraphics[width=0.5\textwidth]{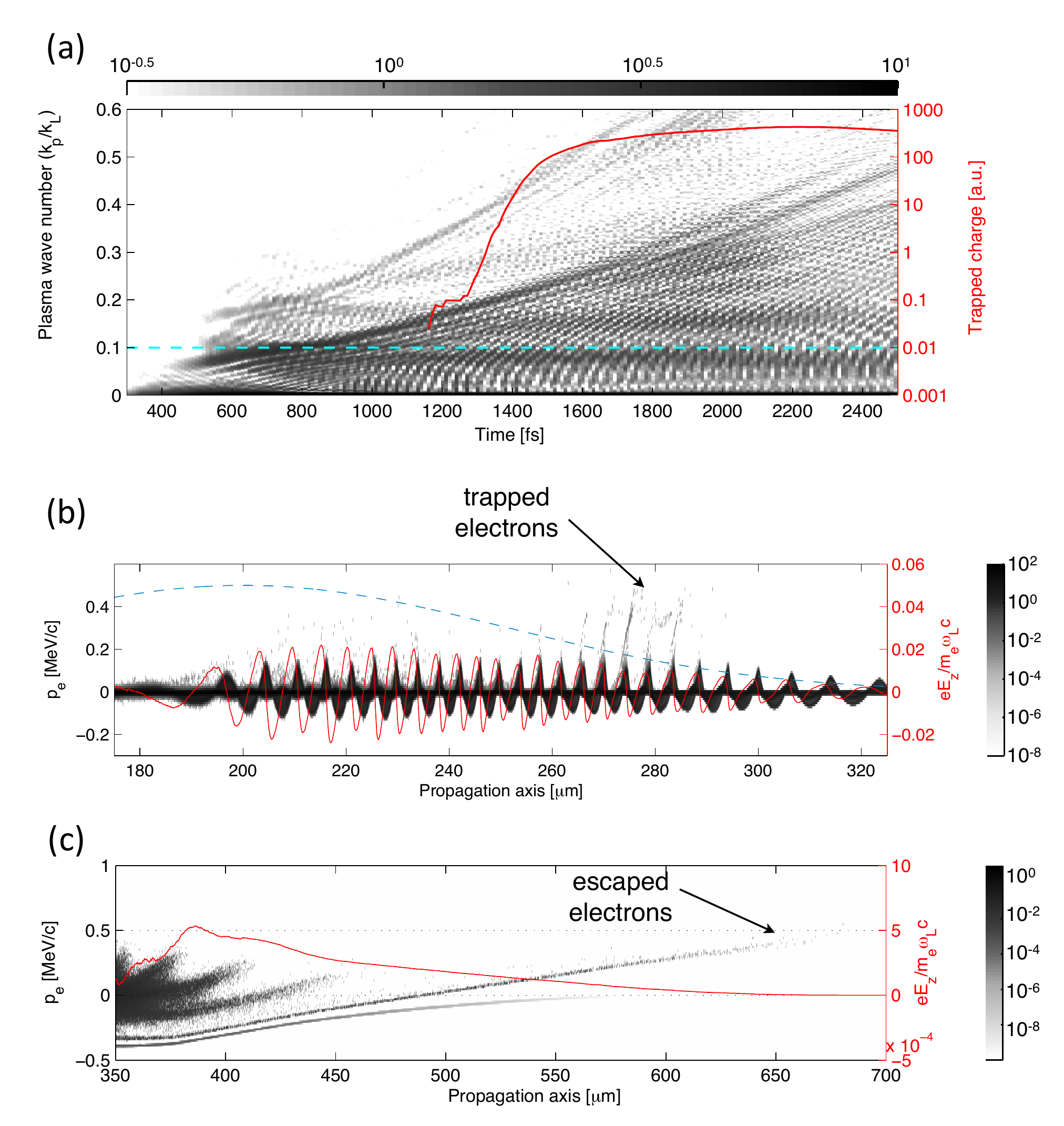}
\caption{(color online). (a) The evolution of plasma wave number spectrum. Solid curve (red) is the total charge of the trapped electrons defined as those with $p_e/m_ec\ge0.4$. (b) Electron \textit{p-x} phase-space plot at $t=1340\unit{fs}$ showing the onset of electron trapping. Dashed blue curve is a lineout of the Gaussian density profile. Solid red curve is the on axis longitudinal electric field. (c) Electron \textit{p-x} phase-space plot at $t=3390\unit{fs}$. Solid curve (red) is the averaged longitudinal electric field in the central axis region. Note the colormaps are in logarithmic scale.}
\label{sim}
\end{center}
\end{figure}

To study the acceleration mechanisms, we performed numerical simulations using the 2D particle-in-cell (PIC) code OSIRIS \cite{osiris}. The simulation ran in a stationary window of 814 $\mu$m $\times$ 38 $\mu$m, with a grid size of $22000\times600$ cells. A Gaussian profile of neutral helium gas was used with the peak centered at 200 $\mu$m [see the lineout in Fig. \ref{sim}(a)]. The peak density was $0.005n_c$, where the plasma critical density $n_c$ is $1.7\times10^{21}\unit{cm^{-3}}$ for $800\unit{nm}$ light. The peak width (FWHM) was 120 $\mu$m as determined from the interferometric measurement (Fig.~\ref{density}). Electrons were produced using the ADK ionization model\cite{ADK}, with 4 particles per cell in each dimension (i.e. 16 fold). The laser parameters were chosen to match our experiment and consisted of a Gaussian spatial profile with a waist of $w_0=1.8\unit{\mu m}$, and a 5th order polynomial temporal profile similar to a Gaussian with a pulse duration of $t_p=32\unit{fs}$. The laser pulse leading edge was initialized at $25\unit{\mu m}$ and focused at $200\unit{\mu m}$. The simulation ran for 3.4 ps. 

The short laser pulse generates large amplitude plasma waves as it propagates through the center of the gas from its ponderomotive force, but not to wavebreaking amplitude. Some time after the laser pulse leaves the plasma, wavebreaking of the plasma waves is observed and electrons are trapped and accelerated. The reason for this trapping is that the plasma waves are formed on the downramp of the gaussian profile, and hence their  phase velocity $v_{ph}$ is not constant in time. In a 1D inhomogeneous plasma, the wavenumber $k$ of a plasma wave varies in time according to  $\partial k_p/\partial t = -\partial \omega_p/\partial x$ \cite{Whitham}. For an appropriately directed traveling wave on a density ramp, $k_p$ increases so that $v_{ph}$ reduces. When the wave phase velocity falls below the maximum electron oscillation velocity in the wake, the charge sheets will cross and trapping (wavebreaking) will commence \cite{Bulanov_PRE_1998}. Electron trapping in a gradual density inhomogeneity can occur several plasma periods behind the laser pulse \cite{Brantov_POP_2008}. This effect is observed in the simulations and illustrated in Fig.~\ref{sim}(a) and (b), where the Fourier transform of the electron density along the central axis is plotted against time, along with the number of trapped electrons. The peak electron density $n_e=0.01n_c$ in this simulation translates to a relativistic plasma  wavenumber $k_p=0.1k_L$ for a linear plasma oscillation, where $k_L=2\pi c/\lambda_L$ is the laser wavenumber in vacuum. 

In Fig.~\ref{sim}(b), we plot the electron phase space showing the electrons injected in the accelerating phase of the wake. At $270\unit{\mu m}$ where trapping occurs, the phase velocity of the plasma wake in the simulation is estimated to be $0.3c$. Using the cold, non-relativistic upper limit for the wave breaking in the one-dimensional approximation $E_{max}=m_e\omega_pv_{ph}/e $~\cite{Dawson1959}, the calculated value is $eE_{max}/m_e\omega_Lc\approx0.02$, which is slightly larger than  the simulated value 0.015 [see Fig.\ref{sim}(b)]. This is due to an equivalent thermal effect in the 2D simulations -- the plasma oscillation has a spread in the fluid speed as seen in Fig.\ref{sim}(b). Equivalent 1D PIC simulations we ran confirm that the trapping condition agrees with the analytical expression very well.

\begin{figure}[t!bp]
\begin{center}
\includegraphics[width=0.4\textwidth]{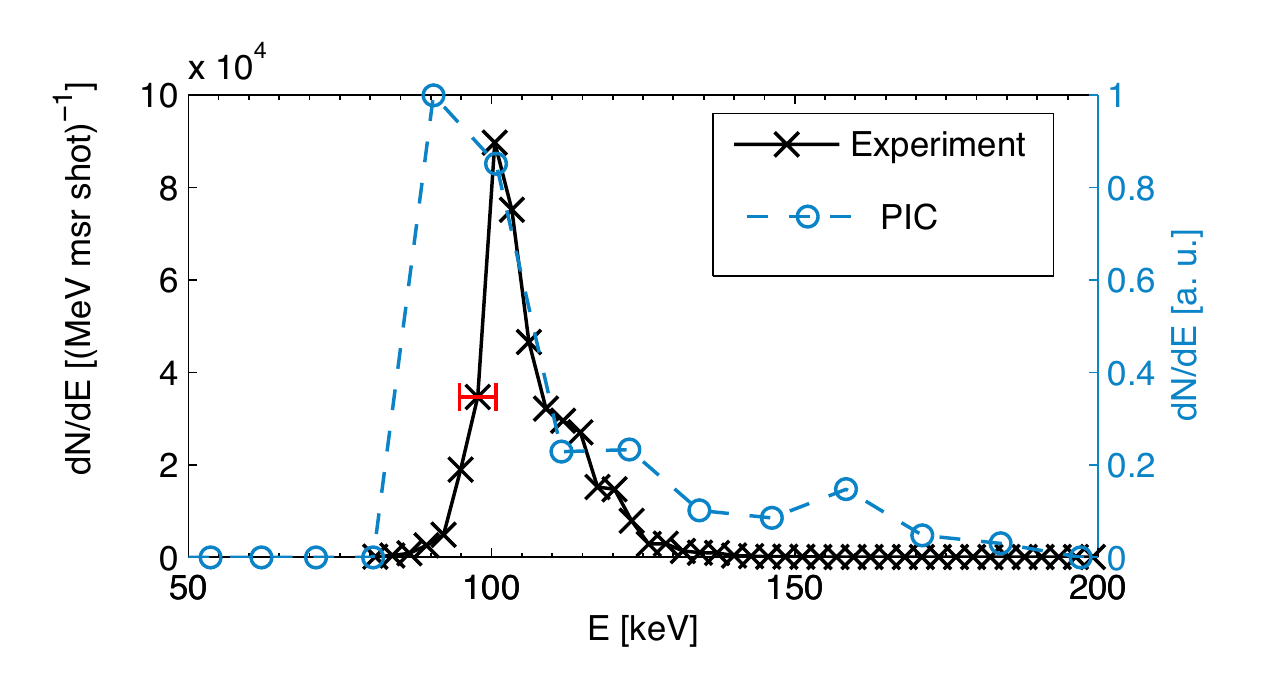}
\caption{(color online). Measured and simulated electron spectra. Black line with crosses: measured spectrum from Fig.~\ref{fig1}(e). Dashed blue line: energy distribution of all electrons in a forward cone angle of 50 mrad obtained from 2D PIC simulations.}
\label{spec}
\end{center}
\end{figure}

These trapped electrons are subsequently accelerated to a sub-relativistic forward momentum in the slow wave. At later times, a longitudinal electric field established by the space charge separation pulls electrons with lower energies back to the plasma but the energetic portion of the trapped electrons escapes as shown in Fig. \ref{sim}(c). To compare the electron spectra in the simulations with the experiment, the energy distribution of all electrons traveling in a forward cone angle of 50 mrad is plotted in the same graph with the measured spectrum in Fig. \ref{spec}. The simulation result shows good agreement within the parameter uncertainty. The gas densities in the simulations were varied over a factor of 4 and $\sim100\unit{keV}$ electrons were observed over the whole density range, with slightly higher energies at lower densities, consistent with the experiments. The simulations suggest that the electron bunch has a few tens of fs duration close to the source.

In conclusion, we have used a high repetition rate 10 mJ short-pulse laser to demonstrate plasma wakefield acceleration of electrons. Reproducible collimated electron beams with a quasi-monoenergetic spectrum in excess of 100 keV can be produced, for our parameters. Simulations show the electrons are trapped and accelerated to sub-relativistic energies in  slow plasma waves. With the capability of operating the system at 500 Hz, further optimization and control of the electron beam profile, energy  and stability can be realized. In addition to demonstrating the scalability of wakefield acceleration to lower energies, such a source may be useful for ultrafast electron diffraction applications.

This work was supported by the NSF/DOE under grant (PHY-09-03557). The authors acknowledge support from Army Research Office (award W911NF11-1-0-116), DNDO and DARPA. The authors would also like to acknowledge the OSIRIS consortium (UCLA/IST Portugal) for the use of the OSIRIS 2.0 framework. Simulations were performed on the Nyx Cluster at the University of Michigan.

\end{document}